\documentclass[12pt]{iopart}
\usepackage{iopams}  
\usepackage{graphicx}
\usepackage{amssymb}
\usepackage{color}
\usepackage{epstopdf}
\usepackage{bm}
\usepackage[normalem]{ulem}

\newcommand{\beq}{\begin{equation}}
\newcommand{\eeq}{\end{equation}}
\renewcommand{\footnoterule}{%
  \kern -3pt
  \hrule width \columnwidth height 1pt
  \kern 2pt
}

\begin{document}

\title{Leading birds by their beaks: the response of flocks to external perturbations}

\author{Nikos Kyriakopoulos and Francesco Ginelli}
\address{SUPA, Institute for Complex Systems and Mathematical Biology, King’s College, University of Aberdeen,
Aberdeen AB24 3UE, United Kingdom}

\author{John Toner}
\address{Department of Physics and Institute of Theoretical Science,
University of Oregon, Eugene, OR 97403, USA}

\date{\today}

\begin{abstract}
We  study  
the asymptotic response  of polar ordered active fluids (``flocks'') to small external  
aligning fields $h$. The longitudinal susceptibility
$\chi_{_\parallel}$ diverges, in the thermodynamic limit, like  $h^{-\nu}$ as
  $h  \rightarrow   0$.  In finite systems  of linear size $L$, $\chi_{_\parallel}$ saturates to a
value $\sim L^\gamma$.
The universal exponents $\nu$ and $\gamma$ depend {\it only } on the
spatial dimensionality $d$, 
and are related to the dynamical exponent $z$ and the
``roughness exponent''  $\alpha$ characterizing the unperturbed flock
dynamics.  Using a well supported conjecture for the values of these two
exponents, we   obtain $\nu   =   2/3$, $\gamma   =   4/5$ in $d   =   2$ and $\nu
  =   1/4$,  $\gamma   =   2/5$ in $d   =   3$. These  values are confirmed by our simulations.
\end{abstract}

\maketitle

\section{Introduction}

Flocking --   the collective motion of many
active particles -- is a ubiquitous emergent
phenomenon that occurs in many
living and synthetic systems 
over a wide range of scales. Examples range from mammal
herds, fish schools and bird flocks to bacteria colonies and cellular migrations, down to 
subcellular  molecular motors
and biopolymers \cite{Ramaswamy2010}. 
Over the last 20 years, studies of minimal models
of self-propelled particles (SPP) \cite{Vicsek95, Vicsek97,
  Chate1, Chate2} and hydrodynamic
continuum theories \cite{TT1, TT2, TT3, TT4, Bertin1, Bertin2, Toner2012} 
have shown 
that the behavior of typical flocking systems 
is essentially determined by
(i) the spontaneous 
breaking of continuous rotational symmetry and (ii) 
the far-from-equilibrium
nature of locally interacting moving particles. 
While the former mechanism is common to many equilibrium systems (ranging from
liquid crystals to magnetic systems and superfluid Helium-4
\cite{PataPok}) which spontaneously align a phase or orientational
degree of freedom, the latter is unique to active matter systems. The
self-propelled motion of active particles results in 
superdiffusive information
propagation 
even in systems without momentum conservation, 
which in turn leads to many striking phenomena never found in equilibrium systems, such as long-range order in two spatial dimensions \cite{TT1}, and anomalously
large number fluctuations \cite{Ramaswamy2003}.

However, little is known concerning the {\it response} of 
moving groups to external
perturbations.
This is an important question in statistical physics:
symmetry breaking  systems are often characterized by their
response to a small external field, and
studying response can also help answer the question of whether a 
generalized fluctuation-dissipation relation (FDR) of some sort
\cite{Vulpiani} holds in
flocks. 
Ethologists, on the other hand,
are interested in response to external threats and more generally in the biological significance of
group response mechanisms. 
Finally,  understanding response is 
essential for {\it controlling} flocking
systems, either biological or artificial.

In equilibrium, the response of systems breaking a
continuous symmetry to a small external field is a classic
problem of statistical field theory, first solved in
\cite{PataPok2}, where it was shown that fluctuations transversal to
order couple to longitudinal ones, yielding a diverging longitudinal
susceptibility in the entire ordered phase.
This is a typical manifestation of symmetry-breaking, and it is a natural
question to wonder how the far-from-equilibrium nature of flocks may change this fundamental result. 

Until now, only a few studies,  mostly numerical, have addressed these
questions. Asymptotic response has been
first studied numerically in the well-known Vicsek model
\cite{Vicsek97}, but that work focused on the behavior of the susceptibility near the transition, rather than in the ordered phase.
Short time response and the dynamic FDR has been investigated numerically in 
the Vicsek model \cite{Chate2} and in the isotropic phase of an active
dumbbells system \cite{Cugliandolo}. The response to finite and/or
localized perturbations, finally, has also been studied in \cite{Ginelli2013, Couzin, Giomi}.

% \sout§{, discussing finite perturbations \cite{Ginelli2013}, response to localized perturbations \cite{Couzin} or investigating dynamic FDR in the isotropic phase of active matter systems\cite{Cugliandolo}.}
Here  we   provide a different approach, combining hydrodynamic theory
results with numerical simulations to characterize the
{\it static} response of ordered flocks to a {\it small} homogeneous external field of
amplitude $h$.  

We  are particularly interested in the asymptotic {\it longitudinal response}
\beq
\chi_{\parallel}     \equiv   \frac{\delta \Phi(h)}{h}\,
\eeq
where $\delta \Phi(h)=\Phi(h)-\Phi(0)$ is the
change in the magnitude of the time-averaged order parameter, which in our case is the mean velocity, due to the applied
field. 
Our main result is the 
scaling law:
\begin{equation}
\chi_{_\parallel}= h^{-\nu}
f
\left(Lh^{1\over z } \right)\propto
\left\{\begin{array}{ll}
%\mbox{h}^{\nu},& L\gg
h^{-\nu},& L\gg
L_c(h) \\ L^{\gamma},&L\ll
L_c(h)
\end{array}\right.
\label{scaling} ,
\end{equation}
where $L_c(h)     \propto      h^{-1/z}$ and, using 
a conjecture  first put forward in
\cite{TT1},
\begin{equation}
\nu={4-d \over d+1} \;,\;\;\;\;\; z = {2 (d + 1) \over 5} \;,\;\;\;\;\; \gamma = {2 (4-d) \over 5}
\label{scaling-exp0}
\end{equation}
for any dimension $3/2   \leq    d   \leq    4$, 
the upper critical dimension. 
In particular, we   have $\nu   =   2/3$, $z   =   6/5$ and 
$\gamma   =    4/5$ in $d  =  2$ and $\nu   =  
1/4$, $z  =    8/5$ and $\gamma   =  2/5$ in $d  =  3$.
For $d  >   4$, on the other hand, we   predict 
$\delta \Phi    \propto   h$.

In the remainder of this paper, we   will first derive the results
  (\ref{scaling}) and (\ref{scaling-exp0}) analytically, and then
  present numerical simulations that confirm them.

\section{Response theory}

We  consider ``dry'' flocks, by which we mean flocks which move over a or through a  {\it static} dissipative substrate or medium that acts as a momentum sink. Total
momentum, thus, is not conserved, and no long ranged hydrodynamic
interactions are present in the system. Obviously,  Galilean 
invariance is broken, since the reference frame in which the static substrate or medium is at rest is preferred.

\subsection{Hydrodynamic description}
The hydrodynamic theory describes flocking by continuous, coarse grained 
number density $\rho({\bf  r}, t)$ and velocity ${\bf v}({\bf r}, t)$ 
fields. 
The hydrodynamic equations of motion governing these fields
in the long-wavelength limit can  be obtained either 
by symmetry arguments \cite{TT1,TT2,TT3,TT4},  or  
by kinetic theory \cite{Bertin1, Bertin2} and describe
  the asymptotic dynamics of polar flocks
regardless of the precise nature of the interactions, provided only that they are  local; in particular, the same hydrodynamic equations apply for both "metric" and "topological" interactions\cite{ST, Topo}.
They are 
\begin{equation}
\partial_t\rho +\nabla\cdot({\bf v}\rho)=0
\label{rho1}
\end{equation}
and, in a schematic notation
\begin{equation}
\partial_{t} {\bf v} + {\bf \Lambda}\left[{\bf \nabla} {\bf v} {\bf v}\right] = U(\rho, |{\bf
  v}|) {\bf v} + {\bf D} \left[{\bf \nabla} {\bf \nabla} {\bf v}\right] +
{\bf F_P} + {\bf f}+{\bf h}
\label{v1}
\end{equation}
where 
\beq
{\bf \Lambda}\left[{\bf \nabla} {\bf v} {\bf v}\right]\equiv\lambda_1({\bf v}\cdot{\bf \nabla}){\bf v}+
\lambda_2({\bf \nabla}\cdot{\bf v}){\bf v}
+\lambda_3 {\bf \nabla}(|{\bf v}|^2)
\label{conv}
\eeq
are all the convective-like
terms permitted by the symmetries and conservation laws of the
system. Here, all three coefficients are, in general, neither zero nor
one, as opposed to systems with Galilean invariance where one has simply
$\lambda_1=1$ and $\lambda_2=\lambda_3=0$ as in usual Navier-Stokes
equations. \\
The viscous terms 
\beq
{\bf D}\left[{\bf \nabla} {\bf \nabla} {\bf v}\right]\equiv D_{1} {\bf \nabla}
({\bf \nabla} \cdot {\bf v})+ 
D_{2}({\bf v}\cdot{\bf \nabla})^{2}{\bf v} + D_{3}\nabla^{2}{\bf v} 
\eeq
reflect the tendency of localized fluctuations in the
velocities to spread out because of local interactions.\\
The pressure term 
\beq
{\bf F_P} \equiv -{\bf \nabla} P_1 -{\bf v} 
\left( {\bf v} \cdot {\bf \nabla}  P_2 \right)
\eeq 
is the sum of an
isotropic and anisotropic pressure terms, the latter being a genuinely
non-equilibrium feature. Both terms tend to suppress local density
fluctuations around the global 
mean value $\rho_0$.
The pressures $P_{1,2}$, and the convective and viscous parameters
$\lambda_k$ and $D_k  > 0$ ($k = 1,2,3$) are 
functions of the
local density $\rho$ and the magnitude
$|{\bf v}|$ of the local velocity.\\
Fluctuations are introduced through a Gaussian white noise ${\bf f}$
with correlations
\begin{equation}
   \left<f_{i}({\bf r},t)f_{j}({\bf r'},t')\right>=\Delta
\delta_{ij}\delta^{d}({\bf r}-{\bf r'})\delta(t-t')
\label{white noise}
\end{equation}
This accounts in a simple way for any source of
  microscopic fluctuations, such as the microscopic noise term opposing order in
  simple SPP models\footnote{One can argue that the fluctuating term
    arising from direct coarse-graining of such models is typically
    multiplicative (i.e., with correlations proportional to the
    density) rather than addictive\cite{ST}. This difference, however, is
  irrelevant for the asymptotic properties discussed here, because the
  {\it local} density fluctuations (not to be confused with the giant number
  fluctuations) in the TT phase are small compared to the
  mean density.}.
%while the
Finally, the local term $U$ simply makes the local $\bf{v}$ have a nonzero
magnitude $v_0(h)$
in the ordered phase. It satisfies the condition $U>0$ for $|{\bf v}|<v_0(h=0)$, $U=0$ for $|{\bf v}|=v_0(0)$, and $U<0$ for $|{\bf v}|>v_0(0)$.
This term thereby spontaneously breaks rotational symmetry 
even in the absence of an external field. Small departures of the statistics of the noise from these assumptions, e.g., slightly  non-Gaussian statistics, or the introduction of ``local color" in the sense of short-ranged spatio-temporal correlations of the noise, change none of the long distance scaling properties of the flock.

Eqs. (\ref{rho1})-(\ref{v1}) are identical to the unperturbed ones
discussed in \cite{Toner2012}, except for the explicit addition of the 
coarse-grained constant field ${\bf h}$ in Eq. (\ref{v1}). By analyticity and
rotational invariance, this field is linearly and isotropically
proportional to the applied microscopic field when those fields are sufficiently small.  

\subsection{Mean-field analysis}
We  first discuss the system in the absence of fluctuations. 
Eqs. (\ref{rho1})-(\ref{v1}) admit a 
spatially uniform steady state solution 
\beq
\begin{array}{l}
\rho({\bf r}, t) = \rho_0\\
{\bf v}({\bf r}, t) = {\bf v}_0({\bf h})
\end{array}
\eeq
For any nonzero external
field let $\hat{\bf e}_{\parallel}$ be the unit vector along ${\bf
  h}   \equiv   h\,\hat{\bf e}_{\parallel}$, while for 
strictly zero field $\hat{\bf e}_{\parallel}$ will be the direction of the
spontaneous symmetry breaking. We  have 
${\bf v}_0({\bf h})  =   v_0(h) \hat{\bf e}_{\parallel}$, with 
the magnitude $v_0(h)$ of the homogeneous velocity ${\bf v}_0({\bf
  h})$ determined by the condition
\beq
U(v_0(h)), \rho_0)\,v_0(h)   +   h   =   0\,.
\eeq
Since $U$ is analytic in $v$, we   have for small fields  
%\begin{equation} v_0(h)=v_0(0) + O(h)\;,
%\label{eq:v0}
%\end{equation} 
%\sout{that is,} 
\beq
v_0(h)   -   v_0(0) \propto h\,,
\label{eq:mf}
\eeq 
where $v_0(0)$ is 
the zero field symmetry broken solution.
It is well known that
sufficiently deep in the ordered phase 
such a zero-field solution is stable 
against spatial perturbations \cite{Bertin2}. In the following we   will
restrict our analysis to this so-called  Toner-Tu (TT) phase.\\
To summarize: in mean field theory, the magnitude of the order parameter 
\beq
\Phi(h)  \equiv   | \langle {\bf v}({\bf r}, t)\rangle |
\label{eq:op}
\eeq 
(here and hereafter $\langle \cdot \rangle$ denotes a global average in space and time) responds linearly in
$h$. 

\subsection{Fluctuations}

We  now move beyond mean field to consider the effect of
fluctuations; we   will show that the corrections to the order 
parameter $\Phi$ due to fluctuations are much larger than linear ones we've just computed at mean field level.

In order to do so, we allow for {\it small} fluctuations around the homogeneous solution, 
\beq
\begin{array}{l}
\rho({\bf r}, t)  =  \rho_0   +   \delta\rho(h; {\bf r}, t)\\ 
{\bf v}({\bf r}, t)   =   {\bf v}_0(h)   +   \delta{\bf v}(h; {\bf r},t)\,,
\end{array}
\eeq
and distinguish between longitudinal and transverse velocity fluctuations, which are respectively parallel ($\parallel$) and perpendicular
($\perp$) to ${\bf v}_0(h)$,
\begin{equation}
\delta{\bf v}(h; {\bf r}, t)
=\delta v_{\parallel} (h; {\bf r}, t)\,{\bf e}_{\parallel} + {\bf
  v}_{\perp}(h; {\bf r}, t)
\label{eq:fluct}
\end{equation}
where we   have made explicit the field dependence of fluctuations. For
simplicity, 
we will hereafter often not explicitly display
the space, time and field dependence of the fluctuations.

Note that, due to number conservation, $\langle\delta \rho \rangle  
=  0$, while
symmetry considerations imply $\langle {\bf v}_{\perp} \rangle = \bf{0}$; that is, 
fluctuations can't steer the global average of $\langle {\bf v}({\bf r}, t) \rangle $ 
away from the external field direction. This implies that corrections to the order parameter
are linear in the longitudinal fluctuations: By making use of
Eqs. (\ref{eq:mf}), (\ref{eq:op}) and (\ref{eq:fluct}) 
we   have
% Guys: I  have replaced the "$\approx$" in the following equation with
% an equality, because I  think the first equality is exact. Let me
% know if you agree.
%I  DO AGREE
\begin{equation}
\Phi(h) = v_0(h) + \langle \delta {\bf v}(h)\rangle = v_0(0) +  \langle \delta v_{\parallel}(h)\rangle  + O(h)
\label{Phi1}
\end{equation}

In order to compute longitudinal fluctuations, we   have to expand the hydrodynamic 
equations (\ref{rho1})-(\ref{v1}) in the small fluctuations $\delta \rho$, $\delta v_{\parallel}$
and ${\bf v}_{\perp}$. We  are interested only in fluctuations that vary slowly in space and time (indeed the
hydrodynamic equations are only valid in this limit), so that space and time derivatives 
of the fluctuations are always of higher order than the fluctuations themselves.
The details of this surprisingly subtle calculation are given in
Appendix A, but (fortunately) they are identical to order $h$ with those for the zero-field case in Ref. \cite{Toner2012}. 
The only difference at $O(h)$ is that
\begin{equation}
U(v_0(h),\rho_0) \approx - \frac{h}{v_0} \neq 0\,.
\label{Uh}
\end{equation}
Because slow modes dominate the long-distance behavior, and, it therefore proves, the small field response, 
we can eliminate the longitudinal fluctuations
from Eqs. (\ref{slow-eq}), since they are a fast mode of the dynamics. The subtle details of this elimination are given in Appendix A; the result is that the longitudinal velocity fluctuation becomes ``enslaved" to the slow modes (that is, its instantaneous value is entirely determined by the instantaneous values of those slow modes)  
via the relation
\begin{equation}
\delta v_{\parallel} \approx  - \frac{|{\bf v}_{\perp} (h) |^2}{2
  v_0(0)} + \mu_1 \delta \rho
+ (\mu_2 \partial_t +
 \mu_3 \partial_{\parallel}) \,\delta \rho + \mu_4  \nabla_{\perp} \cdot {\bf
   v}_{\perp}  + O(h) \,.
\label{eq:parallel}
\end{equation}
Here $\mu_1$ is a constant which depends on the form of $U$. In typical 
flocking models with metric interactions $\mu_1   >   0$ \cite{Bertin2}, so that density fluctuations are 
positively correlated with longitudinal fluctuations at the local level. 
%The further term ``derivatives''  represents 
%terms proportional to one or more spatial and temporal derivatives of
%$\delta \rho$ and ${\bf v}_{\perp}$,
%\beq
% {\rm{derivatives}} = (\mu_2 \partial_t +
% \mu_3 \partial_{\parallel}) \,\delta \rho + \mu_4  \nabla_{\perp} \cdot {\bf
   %v}_{\perp} 
%\label{deriv}
%\eeq
In equation (\ref{eq:parallel}),  $\nabla_{\perp}$ denotes spatial derivatives in the transverse
directions, and the constants $\mu_2$, $\mu_3$ and $\mu_4$ depend on the original parameters of the hydrodynamic equations
(\ref{rho1})-(\ref{v1}). Full details, together with the derivation of
Eq. (\ref{eq:parallel}), can be found in Appendix A, but
the exact form of these constants is unimportant here.
Since these derivative terms are linear in $\delta \rho$ and ${\bf
    v}_{\perp}$, they vanish once averaged over space and time, so
  that from Eq. (\ref{eq:parallel}) we   have
\beq
\langle \delta v_{\parallel} \rangle \approx - \frac{\langle|{\bf v}_{\perp} (h) |^2\rangle}{2 v_0(0)}  + O(h)
\label{mod}
\eeq
which links the global average of transversal and longitudinal
fluctuations and is the analogous of the so-called principle of conservation of
the modulus in an equilibrium ferromagnet \cite{PataPok2}.
From Eq. (\ref{Phi1}) we   finally have 
\begin{equation}
\Phi(h) \approx v_0(0) - \frac{\langle|{\bf v}_{\perp} (h) |^2\rangle}{2 v_0(0)}  + O(h)
\label{Phi2}
\end{equation}
and
\begin{equation}
\delta \Phi(h) \equiv \Phi(h) - \Phi(0) \approx \frac{\langle|{\bf v}_{\perp} (0) |^2\rangle - \langle|{\bf v}_{\perp} (h) |^2\rangle}{2 v_0(0)}  + O(h)
\label{Phi2bis}
\end{equation}

We  are then left with the problem
of determining the fluctuations of the transverse velocity ${\bf v}_{\perp}$ in the presence of a non-zero field $h$. We  will do so by analyzing the equations of motion for ${\bf v}_{\perp}$
and $\delta\rho$, which follow from inserting (\ref{eq:parallel}) into the velocity equation of motion  (\ref{v1})
projected transverse to the direction of mean motion, and into the density equation of motion (\ref{rho1}), and expanding in fields and derivatives. Again, details are relegated to  Appendix A; the result is:
%so that one is left with two coupled equations
%for the true slow modes, $\delta \rho$ (associated with particle number
%conservation) and ${\bf v}_{\perp}$ (associated with continuous symmetry
%breaking), 
\begin{eqnarray}
\label{slow-eq}
\partial_t \delta \rho &=& \left[\,\partial_t \delta \rho\,\right]_{h=0}\\
\partial_t {\bf v}_{\perp} &=& \left[\,\partial_t {\bf v}_{\perp}\right]_{h=0} 
- h_v {\bf v}_{\perp}\nonumber
\end{eqnarray}
where we   have introduced the rescaled field 
\beq
h_v   \equiv   \frac{h}{v_0(0)}
\label{h_v def}
\eeq 
and $\left[\,\partial_t \delta \rho\,\right]_{h=0}$ and $\left[\,\partial_t {\bf v}_{\perp}\right]_{h=0}$
are the terms originally given by Eqs. (2.18) and (2.28)
of \cite{Toner2012} for the zero field case. For
later use, we   denote collectively the parameters appearing in those terms as $\{\mu_i^{(0)}\}$.
While the exact forms of Eqs. (\ref{slow-eq})
are not important for what follows, for completeness we also give them in Appendix A.

\subsection{Renormalization Group}
We  have shown so far that the response $\delta \Phi$ is determined by
the global average of transversal fluctuations, Eq. (\ref{Phi2bis}). 
To compute this quantity, we   proceed by a dynamical renormalization 
group (DRG) analysis \cite{DRG} of Eqs. (\ref{slow-eq}). Once again, this standard analysis is almost identical to
that carried out in \cite{Toner2012} for the zero field case. 
We  start by averaging the equations of motion over the short-wavelength fluctuations: i.e.,   
those with support in the ``shell" of Fourier space $b^{-1} \Lambda  
\le   |{\bf q}_{_\perp} |   \le   \Lambda$,
where $\Lambda$ is an ``ultra-violet cutoff", and $b$ is an arbitrary rescaling factor. 
Then, one  rescales lengths, time,  $\delta\rho$ and ${\bf v}_{_\perp}$ in equations (\ref{slow-eq}) according to 
${\bf v}_{_\perp}  =  b^\alpha  {\bf v}_{_\perp}^{\,\prime}$,
$\delta\rho  =  b^{\alpha}  \delta\rho^{\,\prime}$, 
${\bf r}_{_\perp}  =  b{\bf r}_\perp^{\,\prime}$, $r_{_\parallel}
 =  b^{\zeta} r'_{_\parallel}$, and $t  =  b^zt'$ to restore 
the ultra-violet cutoff  to $\Lambda$ \footnote{One could more generally rescale $\delta \rho$ with a different rescaling
exponent $\alpha_\rho$ from the exponent $\alpha$ used for ${\bf v}_\perp$ . However, since
fluctuations of $\delta \rho$ and ${\bf v}_\perp$ have the same scaling with distance
and time, they prove to rescale with the same exponent $\alpha$ \cite{Toner2012}. Note also that the exponent we   call $\alpha$ here is called $\chi$ in most of the literature; we   have broken this convention here to avoid confusion with the susceptibility $\chi$.}.  The scaling exponents  $\alpha$, $\zeta$, and $z$, 
known respectively as the ``roughness'', ``anisotropy'', and ``dynamical'' exponents, are at this point arbitrary. 

This DRG process leads to a new, ``renormalized''  pair of equations of
motion of the same form as (\ref{slow-eq}), but with ``renormalized''
values  of the parameters, $\{\mu_i^{(0)}\}   \to   \{\mu_i^{(b)}\}$. 
For a suitable choice of the scaling exponents  $\alpha$, $\zeta$, and
$z$, these parameters flow to fixed, finite  limits as
  $b\rightarrow\infty$; that is, 
$\{\mu_i^{(b\rightarrow\infty)}\}\rightarrow\{\mu_i^*\}$; this is referred to as a ``renormalization group fixed point''.
The utility of this choice will be discussed in a moment. 

Since all terms except the $h$ term in Eqs. (\ref{slow-eq}) are
rotation invariant, they can only generate other rotation invariant
terms in the first (averaging) step of the DRG. Hence, they cannot
renormalize $h$, which breaks rotation invariance.
Thus, the only change in the $h$ term in Eqs. (\ref{slow-eq}) occurs in the second (rescaling) step. Since
the coefficient $h_v$ scales as the inverse of time, this 
is easily seen to lead to the recursion relation
\begin{equation} 
h_v = b^{-z} h_v'~~, 
\label{hRR}    
\end{equation}
which -- for the reasons just given -- is exact to linear order in $h$.

By construction, the DRG has the property that correlation functions in the original equations of motions can be 
related to those of the renormalized equations of motion via a simple scaling law. The example of 
interest for our problem is of course the correlation function
\begin{equation}
C\left(L_{\perp},L_{\parallel}, \{\mu_i^0\}, h_v\right)
\equiv \langle | {\bf v}_{\perp} (0)|^2 \rangle - 
\langle | {\bf v}_{\perp} (h)|^2 \rangle 
\end{equation}
Here $L_{\perp}$ and $L_{\parallel}$ are respectively the transverse and longitudinal system size.
The DRG scaling law obeyed by $C$ is thus
\begin{equation}
C\left(L_{\perp},L_{\parallel}, \{\mu_i^0\}, h_v\right)=b^{2 \alpha}
C\left(b^{-1} L_{\perp},b^{-\zeta} L_{\parallel}, \{\mu_i^b\}, b^z h_v\right)
\label{scaling2}
\end{equation}
which follows simply from the fact that $C$ involves two powers of ${\bf v}_{\perp}$, each of which
gives a factor $b^\alpha$. 
%\sout{We  have also made use of the fact that the ``bare'' parameters
%$\{\mu_i^{(0)}\}$ flow to the fixed point $\{\mu_i^*\}$, so that they can be eliminated from the scaling 
%law. Strictly speaking, this is only true in the asymptotic limit
%$b   \to   \infty$, where the rescaled coefficients $\{\mu_i^{(b)}\}$ have 
%fully flown to the fixed point. However, in what follows we   will always 
%choose a value of the rescaling factor $b$ that diverges as the external field 
%amplitude $h$ goes to zero (like $h^{-1/z}$). Hence, our results are 
%asymptotically exact as $h   \to   0$. They will, however, break down for large 
%$h$ for precisely this reason.}

In order to examine the scaling of $C$ with field amplitude $h$,
we use the completely standard\cite{DRG, Ma}
  renormalization group trick of choosing the scaling factor $b$ such that
$b^z h_v $ is equal to some  constant reference field strength
$h_v^*$,  which we  will always choose to have the same value
regardless of the bare value of $h_v$.
This implies that $b = \left({h_v\over
      h_v^*}\right)^{-1/z}$. Note that for small $h$ -- and
    thus small $h_v$ -- this choice implies $b \gg 1$, and that the
    parameters $\{\mu_i^{(b)}\}$ flow to $\{\mu_i^*\}$, their fixed point values. Hence, in the limit of small $h$, the scaling function
  (\ref{scaling2}) can be reduced to
%\left({h\over v_0(0)h_v^*}\right)^{-1/z}$. 
%We  thereby obtain the scaling law
\begin{equation}
C(L_{\perp},L_{\parallel},\{\mu_i^0\},h)=h^{-2 \alpha / z} g(L_{\perp} h^{1/z}, L_{\parallel} h^{\zeta/z})
\label{scaling3}
\end{equation}
where
\beq
g(x,y)\equiv b_0^{2\alpha/z}
C(b_0^{-1/z}x,b_0^{-\zeta/z}y,\{\mu_i^*\},h_v^*)\,\;\;\;\;\mbox{with}\;b_0\equiv  h_v^*\, v_0(0)\,,
\eeq
is a universal scaling function  (since we   always make the same choice of $h_v^*$).

Note that this expression {\it only} applies for small $h$, since it is only in that limit that $b\rightarrow\infty$, and, hence $\{\mu_i^{(b)}\}\rightarrow\{\mu_i^*\}$. Hence, we expect this scaling law to break down for large fields, and, in fact, it does.

We  now focus our attention on roughly square systems, with $L_{\perp}
  \sim   L_{\parallel}   \sim   L$. 
Assuming an anisotropy exponent $0   <   \zeta   <   1$ (as expected
\cite{TT1,TT2,TT3,TT4,Toner2012} for spatial dimensions $d   <   4$), 
we   have for small fields 
\beq
L h^{1/z}   \propto   L_{\perp} h^{1/z}   \ll   L_{\parallel}
h^{\zeta/z}
\eeq
so that 
finite-size scaling is controlled by the transverse flock extension
$L_\perp$ and
we can replace $g$ with the universal scaling function $w(x)
  \equiv   g(x,\infty)$. 
Above the upper critical dimension $d_c   =   4$, where $\zeta   =
  1$ \cite{TT1}, 
scaling is the same in both the transversal and
  longitudinal directions and
we choose instead $w(x)   \equiv   g(x,x)$. 
Doing so, we   finally obtain the scaling law 
\begin{equation}
C(L,h)=h^{1-\nu} \tilde{w}(L h^{1/z})
\label{scaling3bis}
\end{equation}
where \beq
\nu  =  1   +   2 \alpha / z \,.
\eeq
It is now straightforward to relate the order parameter change 
$\delta \Phi(h) $ to this scaling law.
From Eq. (\ref{Phi2bis}) we   have 
\beq
\delta \Phi   \propto   C(L,h)   +  O(h) \,.
\eeq
Note that for $\nu   >   0$ (again a condition we   expect, as discussed below,
to be satisfied for $d   <   4$), we   have $C(L,h)   \gg   O(h)$ 
and the corrections due to fluctuations dominate the mean field ones. 
To lowest order in $h$
\begin{equation}
\delta\Phi = \frac{1}{2 v_0(0)}\left[\langle|{\bf v}_{\perp}(0)|^2\rangle - \langle|{\bf v}_{\perp}(h)|^2\rangle\right] 
= h^{1-\nu} w(L h^{1 / z })
\label{scaling4}
\end{equation}
In the thermodynamic limit, $L h^{1 / z }   \gg   1$ for any non-zero field and 
$w(L h^{1 / z })   \rightarrow   w(\infty)   =   \rm{constant}$, yielding the asymptotic result 
${\delta \Phi\over h}   \propto   h^{-\nu}$. In practice, in any system large enough, the external field suppresses transverse fluctuations,
thus increasing the scalar order parameter according to Eq. (\ref{Phi2}), an effect that below the upper critical dimension  
$d_c  =  4$ proves stronger than mean field corrections linear in $h$. Above
$d_c$, on the other hand, it is known \cite{TT1}  that $\alpha  =  1
- d/2$ and $z  =  2$, which implies   $\nu  =  2 - d/2   \leq   0$; therefore 
corrections due to fluctuations no longer dominate the mean field
ones. Hence, ordinary linear response $\chi_\parallel   \rightarrow
  \rm{constant}$ as $h   \rightarrow   0$ is recovered for $d   >
  4$.

So far, we have kept our discussion of DRG at a qualitative level,
independent of the precise form of the zero-field terms in Eqs. (\ref{slow-eq}).
To be more quantitative for $d   <   4$, we need 
the actual values of the scaling exponents
$\alpha$, $\zeta$, and $z$ for which the DRG flows to a fixed point in
those dimensions. These values actually do depend on the form of
Eqs. (\ref{slow-eq}) (and, in particular, to the nature of their relevant
nonlinear terms), but luckily for small fields $h$ they have to
coincide with their zero-field values. Indeed, there is no reason for
which these zero-field values should be affected by a sufficiently
small rescaled field, $h_v \leq h_v^*$.

In \cite{TT1}, it was argued that for any
dimension $3/2   \leq   d   \leq   4$ these zero-field exponents are 
\begin{equation}
\alpha= {3 - 2 d \over 5}\;,\;\;\;\; z = {2 (d + 1) \over 5} \;,\;\;\;\;\
\zeta= {d + 1 \over 5}.
\label{scaling-exp}
\end{equation}
It has since been since realized \cite{Toner2012} that the original
arguments leading to these values are flawed. However, the simple {\it
  conjecture} that the {\it only} relevant non-linearity 
{\it at the fixed point} 
%\cite{linear}  
is the term proportional to $\lambda_1$ in Eqs. (\ref{v1})-(\ref{conv}) leads to precisely these values for the exponents. 
While this conjecture has
never been proven, there is solid numerical \cite{Chate2, TT2} and even experimental \cite{Rao} evidence supporting the
above scaling exponent values  for  $d  =  2$ and, to a lesser
  extent, $d  =  3$. In the following we will assume this conjecture
  holds, verifying it {\it a-posteriori} by numerical measures of
  asymptotic response in the Vicsek \cite{Vicsek95} model.

Above the upper critical dimension, $d_c  =  4$, finally, 
the scaling exponents take the exact linear values $z  =  2$,
$\zeta  =  1$ and $\alpha  =  1  -   d/2$. 

\subsection{Finite size effects and longitudinal response}
We  conclude this section discussing finite size
effects. The scaling form (\ref{scaling4}) implies that  transverse
fluctuations are suppressed by the field $h$ on length scales  
\beq
L   \gg   L_c (h)   \propto   h^{-1/z} \,.
\eeq
In small systems such that 
$L  \ll   L_c$ (or equivalently for small field $h   \ll   h_c(L)
  \propto   L^{-z}$), 
however, this suppression is ineffective, and leading corrections to the order parameter should revert to linear 
order in the external field $h$. We  can include this behavior in a single universal scaling function $f$ by requiring that 
$f(\infty)   \propto   w(\infty)   =   O(1)$ and $f(x)   \propto
  x^{\nu z}$ for $x\ll 1$. This finally gives, 
\begin{equation}
\delta \Phi = h^{1-\nu}
f\left(h\,L^z \right)\propto
\left\{\begin{array}{ll}
h^{1-\nu},& 
%L\gg L_c(h) 
h \gg L^{-z} \\ 
h L^{\gamma},&
%L\ll L_c(h)
h \ll L^{-z}
\end{array}\right.
\label{scaling5} .
\end{equation}
with $\gamma   =   \nu z$.
This scaling holds for external fields not too large. For $h  >   h_v^*$, on the other hand, the small field approximation discussed 
here is no longer valid, and saturation effects change the scaling (\ref{scaling5}). 
Once expressed in terms of the longitudinal susceptibility 
$\chi_{\parallel}   =   \delta \Phi / h$, our results imply
Eq. (\ref{scaling}), with the scaling exponents given by Eq. (\ref{scaling-exp0}) (according to 
conjecture (\ref{scaling-exp})). 

\section{Numerical simulations}
We  test our predictions (\ref{scaling5}) in two and three dimensions 
by simulating the well known Vicsek model \cite{Vicsek95} 
in an external homogeneous field ${\bf h}$. Each particle -- labeled
by $i   =   1,2,\ldots,N$ -- is defined by a 
position ${\bf r}_i^t$ and a unit direction of motion ${\bf v}_i^t$. The model evolves 
with a synchronous discrete time dynamics
\begin{eqnarray}
{\bf v}_i^{t+1}&=&(\mathcal{R}_\omega \circ \vartheta)\left(\sum_{j \in S_i} {\bf v}_j^t + {\bf h}\right)  \\
{\bf r}_i^{t+1}&=&{\bf r}_i^t + v_m {\bf v}_i^t 
\label{VM}
\end{eqnarray}
where $v_m$ is the particle speed\footnote{Note that $v_m=0$ is the
equilibrium limit of this model \cite{PVM}, which is {\it singular}.},
$\vartheta({\bf w})   =   {\bf w}/|{\bf w}|$ is a normalization operator and 
$\mathcal{R}_\omega$ performs a random rotation (uncorrelated between different times $t$ or particles $i$)
uniformly distributed around the argument vector: $\mathcal{R}_\omega {\bf w}$
is uniformly distributed around ${\bf w}$ inside an arc of amplitude $2\pi   \omega$
(in $d   =  2$) or in a spherical cap spanning a solid angle of
amplitude $4 \pi   \omega$ ($d   =   3$).
The interaction is ``metric'':  that is,  each particle $i$ interacts with all of its neighbors within unit distance.
In the following, we  adopt periodic boundary conditions and choose typical microscopic parameters so that the system lies within the TT phase \cite{Chate2}:
$v_m   =   0.5$,
$\rho_0   =   N/L^d  =   1$ and $\omega  =  0.18$ ($d  =  2$) or $\omega  =  0.11$ ($d  =  3$).
In both dimensions, this choice yields a zero-field order parameter
$\Phi(0)   \approx   0.8$. 
%However, we  have verified that our results also hold
%for different parameter values chosen inside the TT phase and even for
%a topological\cite{Ginelli2010} rather than metric VM\cite{tbw}.

We  perform simulations with different external field amplitudes and with different linear 
system sizes $L$. After discarding a transient $T_0$ sufficiently long for the system to settle 
into the stationary state, we  estimate the mean global order parameter 
\beq
\Phi  =  \left\langle \frac{1}{N} \left |\sum_{i=1}^N {\bf v}_i^t \right|
  \right \rangle_t
\label{op2}
\eeq
and its standard error, given by $S_E   =   \sigma / \sqrt{n}$, with
$\sigma$ being the standard deviation and $n$ the number of independent 
data points. We  estimate $n$ as the total number of stationary points $T$ divided by the 
autocorrelation time $\tau$ of the mean order parameter timeseries, 
$n   =  T/\tau$. In Eq. (\ref{op2}), $\langle \cdot \rangle_t$ denotes
time averages, performed over typically $T=10^6 \sim 10^8$ time
steps. In particular, as the precision of the zero-field order
parameter affects all response and the autocorrelation time decreases with $h$, in our numerical 
simulations we  take care to estimate the zero-field order parameter
over times as large as possible.

\begin{figure}[t!]
\centering
\includegraphics[draft=false,clip=true,width=0.8\textwidth]{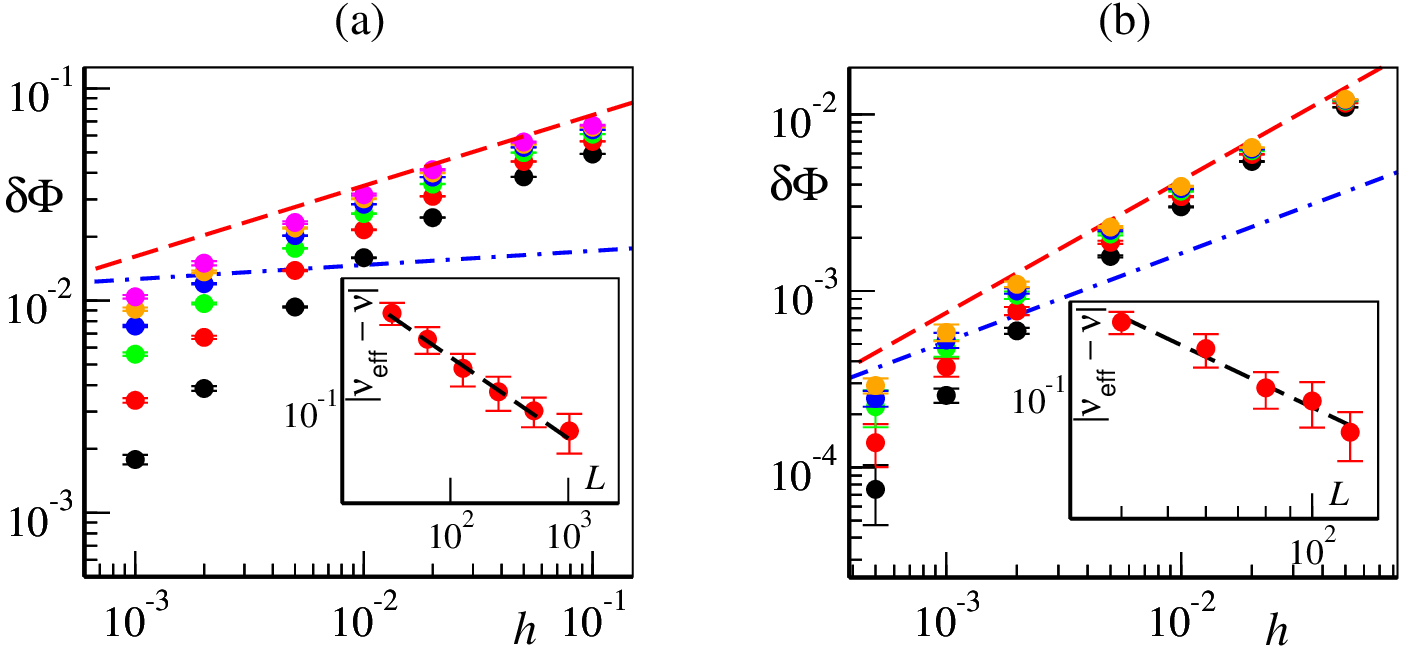}
\caption{(color online) Size-asymptotic regime -- Order parameter change vs. the applied
  field amplitude for different system sizes. a) For $d = 2$, from
  bottom to top, $L = 32, 64, 128, 256,
  512, 1024$. The dashed red line marks the expected asymptotic power law
  behavior $\delta \Phi \sim h^{1/3}$, while the dashed blue line
  marks the upper bound for the $d=2$ equilibrium response $\delta \Phi \sim
  h^{1/15}$. In the inset: Absolute
  difference between the measured effective exponent $\nu_{eff}$ (see
  text) and its expected asymptotic value $\nu$ as a function of
  system size. The dashed black lines marks a power law decay as $1/\sqrt{L}$.
(b) For $d = 3$,  from
  bottom to top, $L = 40, 60, 80, 100, 120$. The dashed red line marks the expected asymptotic power law
  behavior $\delta \Phi \sim h^{3/4}$, while the dashed blue line
  correspond to the $d=3$ equilibrium response $\delta \Phi \sim
  h^{1/2}$. Inset: $d=3$ data as in the inset of panel (a). The dashed black lines marks a power law decay as $L^{-1.5}$.
Error bars measure standard errors (see text). 
All graphs are in a double logarithmic scale.} 
\label{Fig1} 
\end{figure}

We  begin addressing response in the large system size (or large field) regime
$hL^z \gg 1$, where our theory predicts $\delta \Phi \sim
h^{1-\nu}$. Measuring this power law is a particularly difficult task,
as it is sandwiched between saturation effects at larger values of $h$
and the crossover to linear behavior at $h \ll L^{-z}$. We  proceed by
extrapolation, choosing a (somewhat arbitrary) $h$ range of two decades 
and by measuring the effective power law exponent $1-\nu_{eff}$ by linear regression. The 
resulting response $\delta \Phi(h)$ is plotted in Fig.~\ref{Fig1} for
increasing system sizes $L$. As one expects, as system sizes increases,
response curves approach the expected size-asymptotic behavior $\delta \Phi
\sim h^{1-\nu}$. In Fig.~\ref{Fig1}a, for instance, the $d=2$ response
approaches the expected power-law $1-\nu=1/3$. In the inset of
Fig.~\ref{Fig1}a we  further quantify this
convergence plotting $|\nu_{eff}-\nu|$ vs. the system size $L$. This
shows that the effective exponent approaches the predicted one
with corrections of order $1/\sqrt{L}$. We  repeated the same procedure
in $d=3$. As shown in Fig.~\ref{Fig1}b, the approach to the expected
asymptotic exponent $1-\nu=3/4$ is faster, and the difference
$|\nu_{eff}-\nu|$ vanishes faster, as $L^{-1.5}$. In $d=3$ our
simulations are obviously limited to a much smaller range of linear
size values, but it should be noted
that in $d=3$ finite size effects vanish quicker (being the exponent
$z$ larger) while the asymptotic exponent $1-\nu=3/4$ is already quite
close to the value $\delta \Phi \sim h$ expected at low values of
$h$. A faster approach of $\nu_{eff}$ to its asymptotic value is
therefore not completely surprising.

In $d=3$, we  can also easily compare the response behavior with the
equilibrium prediction $\delta \Phi \sim \sqrt{h}$ \cite{PataPok2}
(dashed blue line in Fig.~\ref{Fig1}b). This clearly shows that the
far-from-equilibrium nature of the Vicsek model makes the
susceptibility exponent very different from that in equilibrium ferromagnets. In $d=2$ equilibrium systems with a continuous symmetry
cannot develop long range order,  but, rather, exhibit only a quasi-long range
ordered phase, characterized by scaling exponents that vary continuously with temperature \cite{PataPok, KT}. The equilibrium susceptibility exponent \cite{KT, JKKN} in $d=2$ is given by
\beq
\nu={4-2\eta\over 4-\eta}\,,
\label{nu2dxy}
\eeq
where the order parameter correlation exponent $\eta$ is bounded: $0\le\eta\le1/4$, which implies that the susceptibility exponent $\nu$ varies over an extremely narrow range: $14/15\le\nu\le1$. Our predicted value of $\nu=2/3$ for 2d flocks lies well outside this range; far enough, in fact, that our simulations both support the theory presented here, and rule out any equilibrium interpretation.

\begin{figure}[t!]
\centering
\includegraphics[draft=false,clip=true,width=0.8\textwidth]{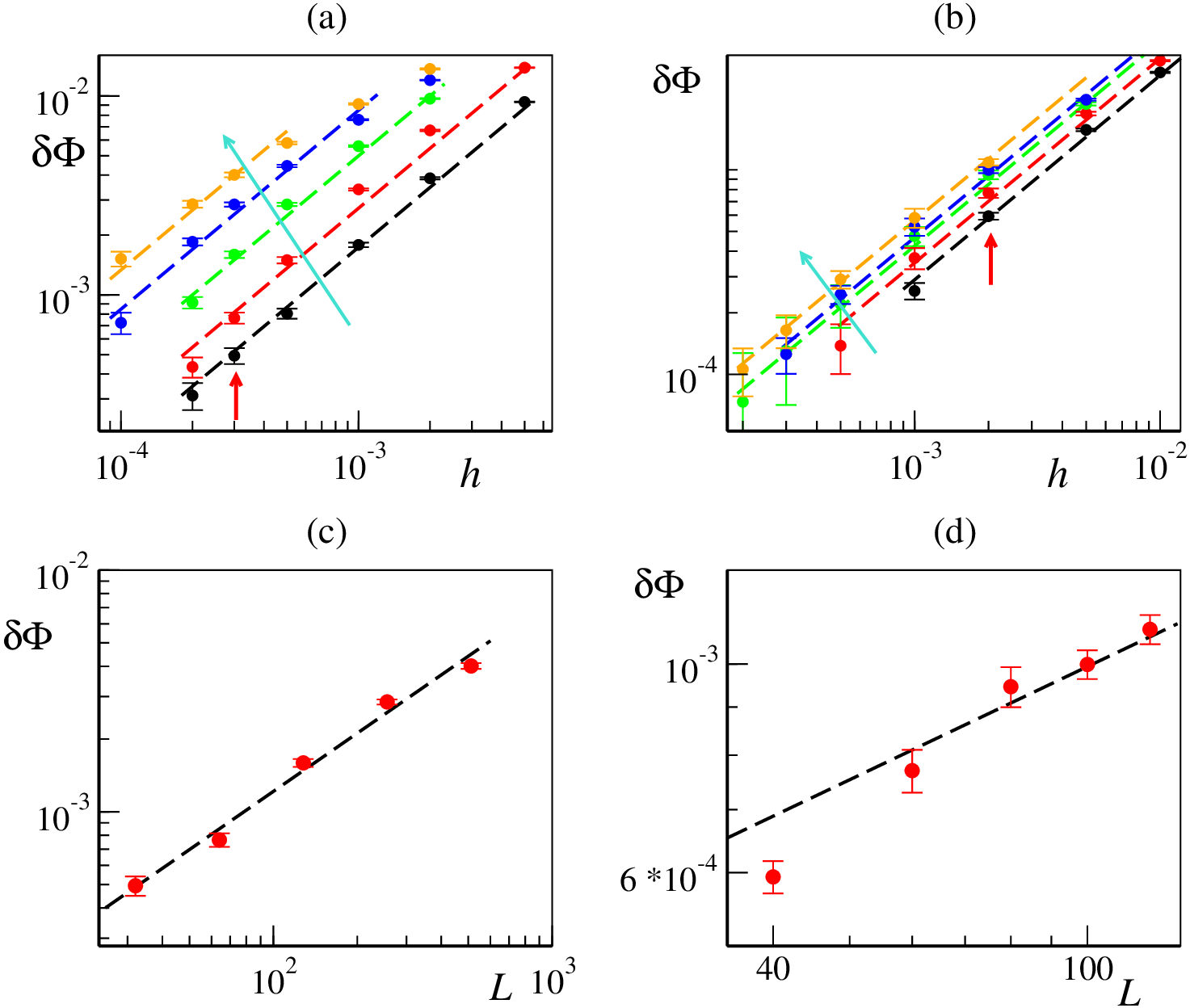}
\caption{(color online) Linear regime. (a)-(b) Order parameter change vs. the applied
  field amplitude in the linear regime for different system sizes (the
  cyan arrow indicates increasing system sizes): a) For $d = 2$ $L = 32, 64, 128, 256,
  512$. (b) For $d = 3$ $L = 40, 60, 80, 100, 120$. The dashed lines
  mark the linear relation $\delta \Phi \sim h$. Error bars measure
  data standard errors (see text).
 (c) $d=2$ Response at fixed $h$ -- as shown by the red arrows in panel
 (a) -- and different system sizes in the linear regime. The dashed
 black line marks a power law with the predicted slope $0.8$. (d) Same
 as in (c), but for $d=3$. The dashed
 black line marks a power law with the predicted slope $0.4$.
All graphs are in a double logarithmic scale.} 
\label{Fig2a} 
\end{figure}

Next, we  consider the linear behavior $\delta\Phi\propto h$ predicted for small system sizes (or small
fields), $hL^z \ll 1$. This inequality imposes a (severe) upper limit on the range of $h$, while there is a lower limit set by our numerical precision in evaluating responses of the order
of $10^{-4}$ or smaller. Nevertheless, our numerical simulations
reveal in both $d=2$ (Fig.~\ref{Fig2a}a) and $d=3$ (Fig.~\ref{Fig2a}b)
a linear growth of the response over more than one decade in the field
amplitude $h$, especially for small system sizes.

\begin{figure}[t!]
\centering
\includegraphics[draft=false,clip=true,width=0.8\textwidth]{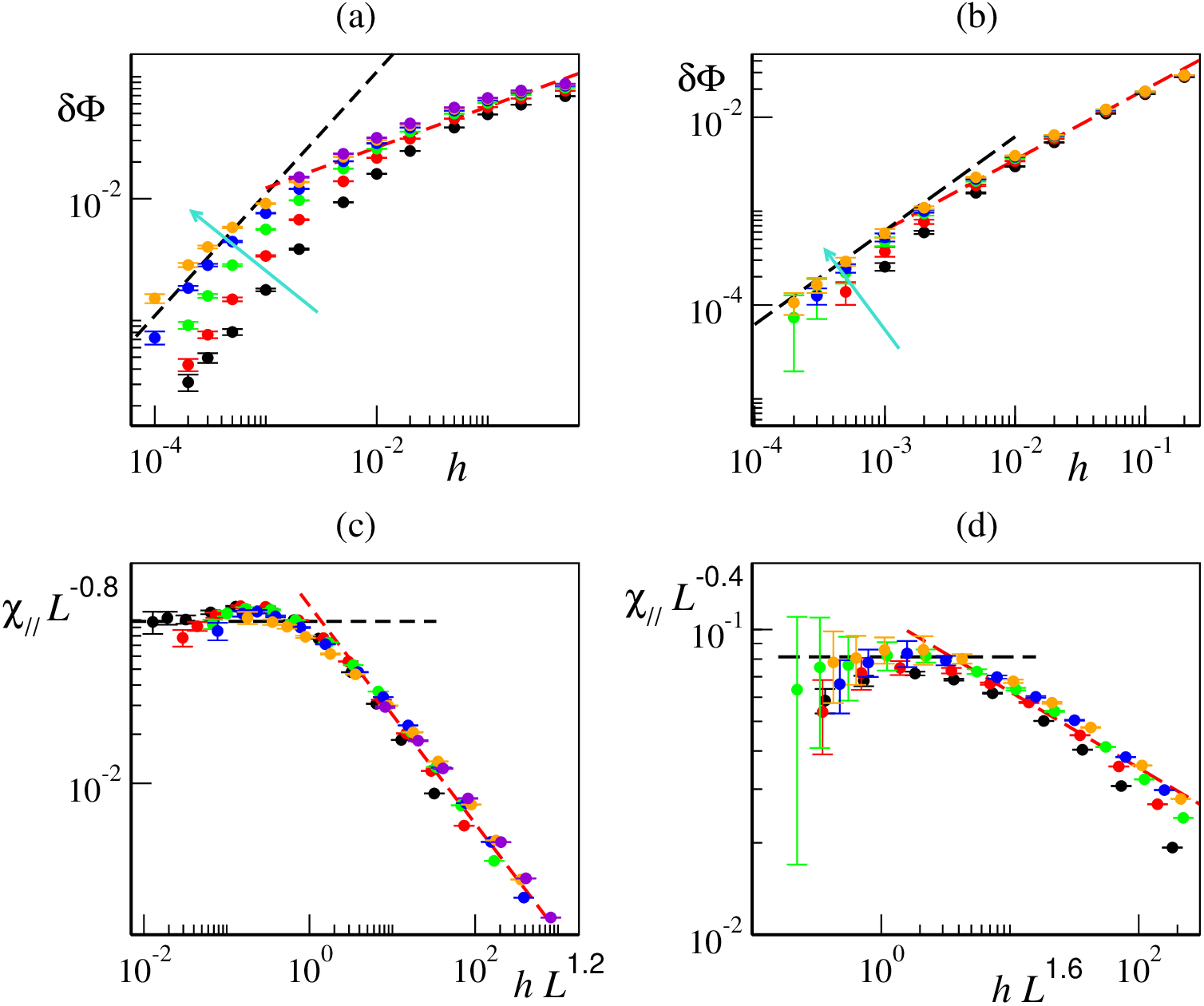}
\caption{(color online) (a)-(b) Order parameter change vs. the applied
  field amplitude in both the linear and size-asymptotic regimes for different system sizes (the arrow indicates
  increasing system sizes): a) For $d = 2$ $L = 32, 64, 128, 256,
  512, 1024$. (b) For $d = 3$ $L = 40, 60, 80, 100, 120$. (c)-(d) data collapse
  for the longitudinal susceptibility according to Eq. (\ref{scaling})
and to the conjectured values for the scaling
exponents (see text). (c) $d = 2$ and (d) $d = 3$. In all panels, the dashed black lines mark
the linear response ($\delta \Phi   \propto   h$ or
$\chi_{\parallel}   \sim L^\gamma$ expected by our
theory for $h   \ll    h_c \sim L^{-z}$. Dashed red lines, on the other hand,
mark the  nonlinear regime predicted for $h   \gg    h_c$, $\delta
\Phi   \propto   h^{1-\nu}$ or $\chi_{\parallel}   \propto  
h^{-\nu}$. We  have $\nu(d = 2) = 2/3$, and $\nu(d = 3) = 1/4$.
Error bars measure standard errors (see text).
All graphs are in a double logarithmic scale.} 
\label{Fig2} 
\end{figure}

By selecting a single $h$ value lying in the linear regime for all
accessible system sizes $L$, one is also able to test the saturation
exponent gamma, $\delta \Phi \sim h\, L^\gamma$. This is done in
Figs.~\ref{Fig2a}c-d, where response values at different linear sizes
$L$ are compared to the predicted power-law with (respectively)
$\gamma = 4/5$ for $d=2$ and $\gamma = 2/5$ in $d=3$. 
We  obtain a good agreement in $d=2$ (the best linear fit being 
$\gamma=0.79(5)$, while data in $d=3$ is less clear, in rough
agreement with the expected power-law behavior only for sizes $L\geq
60$, with a best linear fit of $\gamma=0.48(7)$.

We finally consider the full range of accessible external fields values in Fig.~\ref{Fig2}a-b, which shows data for the accessible
range of external field values in both two and three dimensions. Fields $h$ larger than $h_s \approx 0.1$, of course, are out of the small field regime and show
saturation effects, while due to statistical fluctuations we  have been unable to obtain reliable estimates for 
external fields smaller than $h   \approx   10^{-4}$. Within this
range, comparison with the predicted scaling (\ref{scaling}) (as given
by the dashed lines) is overall satisfactory, especially in $d=2$. By a proper rescaling,
making use of the three scaling exponents (\ref{scaling-exp0}), we  can
also collapse our data at different sizes on roughly a single curve,
as shown in Fig.~\ref{Fig2}c-d.

To summarize, numerical simulations are in good agreement with
our theoretical predictions, at least in $d=2$. Results in $d=3$ are
prone to larger errors and obviously explore a more limited range of
linear sizes, but nevertheless are still compatible with our predictions.

We  also performed a few additional numerical studies of response (not shown here) with different parameter values (but
still in the TT phase), and  in the ordered phase of the so-called
topological Vicsek model \cite{Ginelli2010}, confirming the generality of these results. 

It is also worth commenting on the way the external field is
implemented in the microscopic Vicsek equations (\ref{VM}). 
In Ref. \cite{Chate2} it was argued that different microscopic
implementations could lead to different response, and in particular it
was recommended to choose one by which the external field was
normalized by the local order parameter value, such as in
\beq
{\bf v}_i^{t+1}=(\mathcal{R}_\omega \circ \vartheta)\left[\vartheta\left(\sum_{j \in S_i} {\bf v}_j^t\right) + {\bf h}\right]
\label{v-mod}
\eeq
However, we  do not expect these microscopic details to change the structure of the
hydrodynamic equations (\ref{rho1})-(\ref{v1}), and thus we  do not
expect qualitative differences between the two microscopic external field implementations.
Indeed, our preliminary simulations of Eq. (\ref{v-mod}) (not shown
here) show no qualitative difference from  the response extensively discussed above for Eq. (\ref{VM}).

\section{Conclusions}
So far, we  have only considered the longitudinal
  susceptibility, which characterizes the response of the magnitude of the order
  parameter to a small external field. Simple considerations, based on symmetry, imply that the flock
polarization vector will eventually align with any non-zero
stationary external field, including one applied transversal to the initial
polarization. Thus, for the transvere susceptibility we  have, trivially, 
\beq
\chi_{\perp} \sim \frac{v_0(0)}{h}
\eeq
as in equilibrium systems.

In this paper, we  have fully characterized the static response of homogeneous ordered flocks 
to small external fields for any dimension $d > 3/2$. In particular, below the upper critical dimension
$d_c    =    4$, our results in the thermodynamic limit $L   \to  \infty$
show a diverging longitudinal response for $h   \to   0$, i.e. a diverging susceptibility.
This is ultimately a consequence of the spontaneous symmetry breaking
of the continuous rotation symmetry, albeit the far-from equilibrium nature 
of flocks yields different results from, say, equilibrium ferromagnets
in $d    =    2,3$ \cite{PataPok}.
We  have also fully characterized finite size effects -- typically of great importance in biological 
applications of collective motion -- and verified our results via
numerical simulations.  We believe that the finite numerical values reported in
Ref. \cite{Vicsek97} for the longitudinal susceptibility are entirely due to finite size effects.

Incidentally, our numerics thereby also provide further evidence supporting the conjecture (\ref{scaling-exp}) for the
scaling exponent values \cite{TT1}. 

Our results are expected to hold generically for all collective motion
systems showing a {\it bona fide} TT phase. This class encompasses
both systems with metric interactions and those with topological
interactions. It also
includes the inertial spin model recently put forward in \cite{Attanasi2014}
to account for the turning dynamics measured experimentally in starling flocks. This is because the 
long time hydrodynamic theory of the inertial spin theory relaxes to the TT theory \cite{Cavagna2015}; hence, the static response will be unchanged. Dynamical response (i.e. how quickly the flock turns towards the field direction), 
however, could be different at short times in inertial spin models, while the long-time behavior should be the same.

In future work, we  will explore more thoroughly the phase diagram of Vicsek-like models beyond the TT phase, 
investigating the disordered and phase separated regimes 
%\cite{tbw}.
Other future directions include
the study of the finite-time, dynamical reponse \cite{Cugliandolo97} in both overdamped Vicsek-like models
and inertial spin ones, and the study of spatially localized perturbations.\\

\ack
We  have benefited from discussions with H. Chat\'e and A. Cavagna.
We  acknowledge support from the Marie Curie Career Integration Grant (CIG) PCIG13-GA-2013-618399.
JT also acknowledges support from the SUPA distinguished visitor
program and from the National Science Foundation
through awards \# EF-1137815 and 1006171, and thanks the University of Aberdeen  for their hospitality while this work was underway. FG acknowledges support from 
EPSRC First Grant EP/K018450/1.

%\newpage
\appendix
\section{Expansion of the Hydrodynamic equations for small
fluctuations} 

We first demonstrate that the longitudinal velocity $\delta v_{\parallel}$ is
enslaved to the slow modes $\delta \rho$ and ${\bf v}_{\perp}$.
We follow Ref. \cite{Toner2012} and begin with the hydrodynamic Eqs. (\ref{rho1}) and (\ref{v1}) written out explicitly:
\begin{eqnarray}
\partial_t\rho +\nabla\cdot({\bf v}\rho)=0
\label{conservation}
\end{eqnarray}
\begin{eqnarray}
\partial_{t}
{\bf v}&=&-\lambda_1({\bf v}\cdot{\bf \nabla}){\bf v}-
\lambda_2({\bf \nabla}\cdot{\bf v}){\bf v}
-\lambda_3{\bf \nabla}(|{\bf v}|^2)
+
U(\rho, {\bf v}){\bf v} -{\bf \nabla} P_1 \nonumber\\
&-&{\bf v} 
\left( {\bf v} \cdot{\bf \nabla}  P_2 \right) +
D_{1}{\bf \nabla}
({\bf \nabla}
\cdot {\bf v})+ D_{3}\nabla^{2}{\bf v} +
D_{2}({\bf v}\cdot{\bf \nabla})^{2}{\bf v}+{\bf f} +{\bf h}\,,
\nonumber\\
\label{EOM}
\end{eqnarray}
%
%\begin{eqnarray}
%\partial_{t}{\bf v}&+&\lambda_1 {(\rho,|{\bf v}|)}({\bf v}\cdot{\bf \nabla}){\bf v}+\lambda_2 {(\rho, |{\bf v}|)}({\bf \nabla}\cdot{\bf v}){\bf v}+\lambda_3 {(\rho, |{\bf v}|)}{\bf \nabla}(|{\bf v}|^2)=\alpha{(\rho, |{\bf v}|)}{\bf v}-\beta{(\rho, |{\bf v}|)}|{\bf v}|^{2}{\bf v} -{\bf \nabla} P{(\rho, |{\bf v}|)}  \nonumber \\&-&{\bf v} \left( {\bf v} \cdot{\bf \nabla}  P_2 {(\rho, |{\bf v}|)}\right) +D_{L}{(\rho, |{\bf v}|)}{\bf \nabla}({\bf \nabla}\cdot {\bf v})+ D_{3}{(\rho, |{\bf v}|)}\nabla^{2}{\bf v} +D_{2}{(\rho, |{\bf v}|)}({\bf v}\cdot{\bf \nabla})^{2}{\bf v}+{\bf f}\label{EOM}
%\end{eqnarray}
%
where, as noted in the main text,  the parameters $\lambda_i  (i  = 1 \to 3)$,
the local term $U$, the  ``isotropic Pressure'' $P(\rho,
|{\bf v}|)$ and the  ``anisotropic Pressure''$P_2 (\rho, |{\bf v}|)$
are, in general, functions of the density $\rho$ and the magnitude
$|{\bf v}|$ of the local velocity. It is useful to Taylor expand $P_1$ and 
$P_2$ around the equilibrium density $\rho_0$:
\begin{eqnarray}
P_1=\sum_{n=1}^{\infty} \sigma_n ( |{\bf v}|)
(\rho-\rho_0)^n
\label{P rho}
\end{eqnarray}
\begin{eqnarray}
P_2=P_2(\rho, |{\bf v}|)=\sum_{n=1}^{\infty} \Upsilon_n ( |{\bf v}|)(\rho-\rho_0)^n~~.
\label{P2 rho}
\end{eqnarray}
Here  $D_{1}$, $D_{2}$ and $D_{3}$ are all
positive in
the ordered state.  \\
Again as discussed in the main text,
in the ordered phase,   the velocity field can be written as:
\begin{eqnarray}
   {\bf v}=v_{0}{\bf{e}}_{{_\parallel}}+{\bf \delta v}= (v_{0}+\delta v_{_\parallel}){\bf e}_{{_\parallel}}+{\bf  v}_\perp~~
\label{v fluc}
\end{eqnarray}
(for simplicity, here and hereafter, we write $v_0 \equiv v_0(h)$).\\
%   where
% $v_{0}{\bf e}_{{_\parallel}}=<{\bf v}>$ is the spontaneous average
%value of
%${\bf v}$ in the ordered phase, and the fluctuations $\delta v_{_\parallel}$ and $\vec{ v}_\perp$ of ${\bf v}$ about this mean velocity along and perpendicular to the direction of the mean velocity are assumed to be small. Indeed, we   will be shortly be expanding the equation of motion (\ref{EOM}) in these quantities. (We  will also hereafter be using the subscripts ${_\parallel}$ and $\perp$ to denote components of any vector along and perpendicular to the mean velocity $<{\bf v}>$.)
Taking the dot product of both sides of equation (\ref{EOM}) with ${\bf v}$ itself, we   obtain:
%+\lambda_2 {(\rho, |{\bf v}|)}({\bf \nabla}\cdot{\bf v}){\bf v}+\lambda_3 {(\rho, |{\bf v}|)}{\bf \nabla}(|{\bf v}|^2)=\alpha{\bf v}-\beta|{\bf v}|^{2}{\bf v} -{\bf \nabla} P -{\bf v} 
%\left( {\bf v} \cdot{\bf \nabla}  P_2 \right) +D_{L}{\bf \nabla}
%({\bf \nabla}\cdot {\bf v}) \nonumber \\&+& D_{3}\nabla^{2}{\bf v} +D_{2}({\bf v}\cdot{\bf \nabla})^{2}{\bf v}+{\bf f}

\begin{eqnarray}
{1\over 2}\left(\partial_{t}|{\bf v}|^2\right)&+&{1\over
  2}\left(\lambda_1 + 2 \lambda_3)({\bf v}\cdot{\bf \nabla})|{\bf
    v}|^2\right) + \lambda_2({\bf \nabla}\cdot{\bf v})|{\bf v}|^2 =
U(|{\bf v})|{\bf v}|^{2} \nonumber\\
&-&{\bf v} \cdot{\bf \nabla}  P_1
-|{\bf v}|^{2}{\bf v} \cdot{\bf \nabla}  P_2 +D_{1} {\bf v}\cdot{\bf \nabla}
({\bf \nabla}\cdot {\bf v}) 
+ D_{3}{\bf v}\cdot\nabla^{2}{\bf v} \nonumber\\&+&
D_{2}{\bf v}\cdot\left(({\bf v}\cdot{\bf \nabla})^{2}{\bf v}\right)
+{\bf v}\cdot{\bf f}+{\bf v}\cdot{\bf h}~.
\nonumber\\
\label{v parallel elim}
\end{eqnarray}

In this hydrodynamic approach, 
we   are interested only in fluctuations ${\bf \delta v}({\bf r}, t)$ and $\delta \rho({\bf r}, t)$
that vary slowly in space and time. (Indeed, the hydrodynamic equations (\ref{EOM}) and (\ref{conservation}) are only valid in this limit). Hence, terms involving space and time derivatives of 
${\bf \delta v}({\bf r}, t)$ and $\delta \rho({\bf r}, t)$
are always negligible, in the hydrodynamic limit, compared to terms involving the same number of powers of fields without any time or space derivatives.

Furthermore, the fluctuations 
${\bf \delta v}({\bf r}, t)$ and $\delta \rho({\bf r}, t)$ can
themselves be shown to be small in the long-wavelength limit \cite{Toner2012}. Hence, we   need only keep terms in equation (\ref{v parallel elim}) up to linear order in 
${\bf \delta v}({\bf r}, t)$ and $\delta \rho({\bf r}, t)$. The 
${\bf v}\cdot{\bf f}$ term can likewise be dropped, since it only leads to a term of order 
${\bf v}_{_\perp} f_{_\parallel}$ in the ${\bf v}_{_\perp}$ equation of motion, which is negligible (since ${\bf v}_{_\perp}$ is small) relative to the ${\bf f}_\perp$ term already there. 

In addition, treating the magnitude $h$ of the applied field as a small quantity, we need only keep terms involving $h$ that  are proportional to $h$ and {\it independent} of the small fluctuating quantities ${\bf \delta v}$ and $\delta\rho$.

These observations can be used to eliminate many of the terms in equation (\ref{v parallel elim}), and solve for the quantity $U$;
%\begin{eqnarray}
%U \equiv  \alpha(\rho , |{\bf v}|)-\beta (\rho ,
%|{\bf v}|)|{\bf v}|^2 ~;
%\label{Udef}
%\end{eqnarray}
the  solution  is:
\begin{eqnarray}
U=-{h\over v_0}+\lambda_2{\bf \nabla}\cdot{\bf v}+{\bf v}\cdot{\bf
  \nabla}P_2+{\sigma_1\over v_0}\partial_{_\parallel} \delta
\rho+{1\over 2 v_0}\left(\partial_t + \lambda_4 \partial_{_\parallel}
\right)\delta v_{_\parallel} , 
%\nonumber\\
\label{Usol}
\end{eqnarray}
where we've defined 
\begin{eqnarray}
\lambda_4\equiv(\lambda_1+2\lambda_3)v_0~.
\label{gamma_2def}
\end{eqnarray}

We can now express the longitudinal velocity $\delta v_{\parallel}$ in
terms of the slow modes using equation (\ref{Usol}) and 
the expansion 
\begin{eqnarray}
U\approx-\Gamma_1\left(\delta v_{_\parallel} +{|{\bf v}_{_\perp}|^2\over 2 v_0}\right) - \Gamma_2 \delta \rho ~~,
\label{Uexp}
\end{eqnarray}
where 
we've defined 
\begin{eqnarray}
\Gamma_1 \equiv -\left({\partial U
 \over \partial |{\bf v}|}\right)^0_{\rho} &,
&\Gamma_2 \equiv - \left({\partial U
 \over \partial \rho}\right)^0_{|{\bf v}|}~,
\label{gamma12 def}
\end{eqnarray}
with, here and hereafter, super- or sub-scripts $0$ denoting functions
of $\rho$ and $|{\bf
  v}|$ evaluated at $\rho = \rho_0$ and $ |{\bf
  v}|=v_0$. 
We've also used the expansion
(\ref{v fluc}) for the velocity in terms of the fluctuations $\delta v_{_\parallel}$ and $\vec{ v}_\perp$ to write
\begin{eqnarray}
|{\bf v}|=v_0+\delta v_{_\parallel} +{|{\bf v}_{_\perp}|^2\over 2 v_0}+O(\delta v_{_\parallel} ^2, |{\bf v}_{_\perp}|^4)~,
\label{speed}
\end{eqnarray}
and kept only terms that an DRG analysis shows to be relevant in the
long wavelength limit \cite{Toner2012}.
Inserting  (\ref{Uexp}) into (\ref{Usol})  gives:

\begin{eqnarray}
-\Gamma_1\left(\delta v_{_\parallel} +{|{\bf v}_{_\perp}|^2\over 2 v_0}\right) - \Gamma_2 \delta \rho &=&-{h\over v_0}+\lambda_2{\bf \nabla}_{_\perp}\cdot{\bf v}_{_\perp}+\lambda_2\partial_{_\parallel} \delta v_{_\parallel}\nonumber\\&+&{(\Upsilon_1 v_0^2+\sigma_1)\over v_0}\partial_{_\parallel} \delta \rho+{1\over 2 v_0}\left(\partial_t + \lambda_4 \partial_{_\parallel} \right)\delta v_{_\parallel}~~,\nonumber\\
\label{v par 1}
\end{eqnarray}
where we've kept only linear terms on the right hand side of this equation, since the non-linear terms are at least of order derivatives of $|{\bf v}_{_\perp}|^2$, and hence negligible, in the hydrodynamic limit, relative to the  $|{\bf v}_{_\perp}|^2$ term explicitly displayed on the left-hand side.

This equation can be solved iteratively for $\delta v_{_\parallel}$ in
terms of ${\bf v}_{_\perp}$, $\delta \rho$, and its derivatives. To
lowest (zeroth) order in derivatives, 
$\delta v_{_\parallel} \approx \mu_1 \delta\rho +
\frac{h}{v_0 \Gamma_1 } $
where we have defined 
\beq
\mu_1=-{\Gamma_2\over \Gamma_1}\,.
\label{mu1}
\eeq
Inserting this approximate expression for $\delta v_{_\parallel} $ into equation (\ref{v par 1}) everywhere  $\delta v_{_\parallel}$ appears on the right hand side of that equation gives $\delta v_{_\parallel}$ to first order in derivatives:
\begin{eqnarray}
\delta v_{_\parallel}\approx -{|{\bf v}_{_\perp}|^2\over 2 v_0}+\mu_1\,\delta\rho
+{\Gamma_2\over v_0\Gamma_1^2}\partial_t \delta \rho-{\lambda_5
\over\Gamma_1} \partial_{_\parallel} \delta\rho -
{\lambda_2\over\Gamma_1}{\bf \nabla}_{_\perp}\cdot{\bf
  v}_{_\perp}+{h\over v_0\Gamma_1}~,
%\nonumber\\
\label{v par 2}
\end{eqnarray}
where we've defined
\begin{eqnarray}
\lambda_5 \equiv {(\Upsilon_1v_0^2+\sigma_1)\over v_0} - 
{\Gamma_2\over\Gamma_1}\left(\lambda_2+{\lambda_4\over v_0}
\right)= {(\Upsilon_1v_0^2+\sigma_1)\over v_0} + \mu_1 
\left(\lambda_1+\lambda_2+2\lambda_3
\right)~.\nonumber\\
\label{lambda_5 def}
\end{eqnarray}
(In deriving the second equality in (\ref{lambda_5 def}), we've used the definition (\ref{gamma_2def}) of $\lambda_4$.)\\
Eq. (\ref{v par 2}) coincides with Eq. (\ref{eq:parallel}) in the main
text with
\begin{eqnarray}
\mu_2 &=& {\Gamma_2\over v_0\Gamma_1^2}\\
\mu_3 &=& -{\lambda_5\over\Gamma_1} \\
\mu_4 &=& - {\lambda_2\over\Gamma_1}
\label{mu2-4}
\end{eqnarray}
and expresses the enslaving of
$\delta v_{\parallel}$ to the slow modes $\delta \rho$ and ${\bf v}_{\perp}$.\\

We finally derive the equations of motion for the slow modes.
Inserting the expression (\ref{Usol}) for $U$ back into equation (\ref{EOM}), %( where $U$ appears by virtue of its definition(\ref{Udef})), 
we   find that $P_2$ and 
$\lambda_2$ cancel out of the 
${\bf v}$ equation of motion, leaving
\begin{eqnarray}
\partial_{t} {\bf v} &=&-\lambda_1({\bf v}\cdot{\bf \nabla}){\bf v}-\lambda_3{\bf \nabla}(|{\bf v}|^2)
-{h\over v_0}{\bf v}+{\sigma_1\over v_0}
{\bf v} (\partial_{_\parallel} \delta \rho)-{\bf \nabla} P_1 + D_{1}{\bf \nabla}
({\bf \nabla}
\cdot {\bf v}) 
\nonumber\\
&+& D_{3}\nabla^{2}{\bf v}+
D_{2}({\bf v}\cdot{\bf \nabla})^{2}{\bf v}+\left[{1\over 2 v_0}\left(\partial_t + \lambda_4 \partial_{_\parallel} \right)\delta v_{_\parallel}
\right]{\bf v}+{\bf f}+{\bf h}~~.
\nonumber\\
\label{EOM2}
\end{eqnarray}

This can be made into an equation of motion for ${\bf v}_{_\perp}$
involving only ${\bf v}_{_\perp}({\bf r}, t)$ and $\delta \rho({\bf
  r}, t)$ by i) projecting (\ref{EOM2}) perpendicular to the direction of mean flock
motion ${\bf e}_{_\parallel}$, and ii) eliminating $\delta v_{_\parallel}$ by inserting equation (\ref{v par 2}) 
into the equation of motion (\ref{EOM2}) for
${\bf v}$.
Using the expansions
(\ref{v fluc}), (\ref{speed}) and neglecting ``irrelevant'' terms we have:

% and projecting that equation perpendicular to the mean direction of flock motion ${\bf e}_{_\parallel}$ %and (\ref{conservation}) for $\delta\rho$,  
%gives, neglecting ``irrelevant'' terms:

\begin{eqnarray}
\partial_{t} {\bf v}_{_\perp} &=& -\lambda_1^0 v_0 \partial_{{_\parallel}} 
{\bf v}_{_\perp} -\lambda_1^0 \left({\bf v}_{_\perp} \cdot
{\bf \nabla}_{_\perp}\right) {\bf v}_{_\perp} -g_1\delta\rho\partial_{{_\parallel}} 
{\bf v}_{_\perp}-g_2{\bf v}_{_\perp}\partial_{{_\parallel}}
\delta\rho 
-{c_0^2\over\rho_0}{\bf \nabla}_{_\perp}
\delta\rho \nonumber\\
&-&g_3{\bf \nabla}_{_\perp}(\delta \rho^2)+
D_B{\bf \nabla}_{_\perp}\left({\bf \nabla}_{_\perp}\cdot{\bf v}_{_\perp}\right)+
D_3\nabla^{2}_{_\perp}{\bf v}_{_\perp} +
D_{{_\parallel}}\partial^{2}_{{_\parallel}}{\bf v}_{_\perp}
+g_t\partial_t{\bf
  \nabla}_{_\perp}\delta\rho \nonumber\\
&+&g_{_\parallel}\partial_{_\parallel}{\bf
  \nabla}_{_\perp}\delta\rho+{\bf f}_{\perp} 
-h_v{\bf v}_{_\perp}\nonumber\\
\label{vEOMbroken}
\end{eqnarray}

where  we've defined $h_v$ as in the main text, and
\begin{eqnarray}
D_B\equiv D_1+{2v_0\lambda_3^0\lambda_2^0\over\Gamma_1}~, 
\label{DBeff def}
\end{eqnarray}
\begin{eqnarray}
D_{{_\parallel}}\equiv D_{3}+D_{2}v_0^2~,
\label{Dpar def}
\end{eqnarray}
\begin{eqnarray}
g_{1}\equiv \mu_1 \lambda_1^0 + v_0\left({\partial\lambda_1 \over \partial \rho}\right)_0 ~,
%-{\Gamma_2\lambda_1\over\Gamma_1}  
\label{g_1 def}
\end{eqnarray}
\begin{eqnarray}
g_2 \equiv -\mu_1 {\lambda_4^0\over v_0} -{\sigma_1^0\over v_0}~, 
%{\Gamma_2\lambda_4\over\Gamma_1v_0}
\label{g_2  def}
\end{eqnarray}
%\begin{eqnarray}
%\label{
%$c_0^2$,  and % def}
%\end{eqnarray}
\begin{eqnarray}
g_3 \equiv\sigma_2^0+\mu_1^2 \lambda_3^0+  v_0\mu_1 \left({\partial\lambda_3
\over \partial \rho}\right)_0~,
\label{g_3  def}
\end{eqnarray}
\begin{eqnarray}
 c_0^2 \equiv \rho_0\sigma_1^0 + 2\rho_0v_0\lambda_3^0\mu_1~, 
 \label{c0def}
 \end{eqnarray}
 
\begin{eqnarray}
g_{t}
%$ are all constants expressible in terms of the parameters in equation (\ref{v par 2}).
\equiv 2 \mu_1 \lambda_3^0 
%-{2\Gamma_2\lambda_3\over \Gamma_1^2}~,
\label{nu_t def}
\end{eqnarray}
and
\begin{eqnarray}
g_{_\parallel}\equiv {2v_0\lambda_3^0\lambda_5^0\over \Gamma_1} -
\mu_1 D_1 ~.
%+{\Gamma_2D_1\over\Gamma_1}~.
\label{nu par def}
\end{eqnarray}
\\Finally, using  (\ref{v fluc}), (\ref{speed}) and (\ref{v par 2}) in the equation of motion (\ref{conservation}) for $\rho$ gives, again neglecting irrelevant terms:

\begin{eqnarray}
\partial_t\delta \rho &=& -\rho_0{\bf \nabla}_{_\perp}\cdot{\bf v}_{_\perp}
-w_1{\bf \nabla}_{_\perp}\cdot({\bf v}_{_\perp}\delta\rho)-v_2
\partial_{{_\parallel}}\delta
\rho +
D_{\rho{_\parallel}}\partial^2_{_\parallel}\delta\rho+D_{\rho{_\perp}}\nabla^2_{_\perp}\delta\rho\nonumber\\
&+&D_{\rho v} \partial_{{_\parallel}}
\left({\bf \nabla}_{_\perp} \cdot {\bf v}_{_\perp}\right)+\rho_0
\mu_2 \partial_t\partial_{_\parallel}\delta\rho
-\mu_1\partial_{_\parallel}(\delta\rho^2)+\mu_5\partial_{_\parallel}(|
{\bf v}_{_\perp}|^2)~, \nonumber \\
\label{cons broken}
\end{eqnarray}

where we've defined:
\begin{eqnarray}
v_2\equiv v_0 + \mu_1 \rho_0
%- {\rho_0\Gamma_2\over \Gamma_1},~
\label{v_2 def1}
\end{eqnarray}
\begin{eqnarray}
\mu_5 \equiv {\rho_0\over 2v_0}~ ,
\label{w_3 def}
\end{eqnarray}
\begin{eqnarray}
D_{\rho{_\parallel}}&\equiv&{\rho_0\lambda_5^0\over\Gamma_1}=-\rho_0
\,\mu_3
%\nonumber\\
%&= &{\rho_0\over \Gamma_1}\left({(\Upsilon_1v_0^2+\sigma_1)\over v_0}-{\Gamma_2\over \Gamma_1} \left(\lambda_1 + \lambda_2 + 2\lambda_3 \right) \right),
%\nonumber \\
\label{Drhopardef}
\end{eqnarray}
and, last but by no means least,
\begin{eqnarray}
D_{\rho v} \equiv {\lambda_2^0\rho_0\over \Gamma_1} = -\rho_0 \,\mu_4, 
\label{D_rho v def}
\end{eqnarray}

The parameter $D_{\rho{_\perp}}$ is actually zero at this point in the calculation, but we've included it in equation (\ref{cons broken}) anyway, because it is generated by the nonlinear terms under the Renormalization Group. Likewise, the parameter $w_1=1$, but will change from that value upon renormalization.

The equation of motion (\ref{cons broken}) is, as claimed in the main text, exactly the same as that in the absence of the external field $h$, while the equation of motion (\ref{vEOMbroken}) is
of the form (\ref{slow-eq}), with 
\begin{eqnarray}
 \left[\,\partial_t {\bf v}_{\perp}\right]_{h=0}&\equiv&  -\lambda_1^0 v_0 \partial_{{_\parallel}} 
{\bf v}_{_\perp} -\lambda_1^0 \left({\bf v}_{_\perp} \cdot
{\bf \nabla}_{_\perp}\right) {\bf v}_{_\perp} -g_1\delta\rho\partial_{{_\parallel}} 
{\bf v}_{_\perp}-g_2{\bf v}_{_\perp}\partial_{{_\parallel}}
\delta\rho \nonumber\\
&-&{c_0^2\over\rho_0}{\bf \nabla}_{_\perp}
\delta\rho 
-g_3{\bf \nabla}_{_\perp}(\delta \rho^2)+
D_B{\bf \nabla}_{_\perp}\left({\bf \nabla}_{_\perp}\cdot{\bf v}_{_\perp}\right)+
D_3\nabla^{2}_{_\perp}{\bf v}_{_\perp} \nonumber\\
&+&
D_{{_\parallel}}\partial^{2}_{{_\parallel}}{\bf v}_{_\perp}
+g_t\partial_t{\bf
  \nabla}_{_\perp}\delta\rho 
+g_{_\parallel}\partial_{_\parallel}{\bf
  \nabla}_{_\perp}\delta\rho+{\bf f}_{\perp} 
\nonumber\\
\end{eqnarray}

%\newpage

\end{document}